\documentclass[twocolumn,amsmath,showkeys,amssymb,superscriptaddress,longbibliography,floatfix, prl]{revtex4-2}
\usepackage{graphicx, subfigure}
\usepackage{amssymb, amsmath,amssymb,amsfonts}
\usepackage{amsthm,mathrsfs,amsopn}
\usepackage{dcolumn}
\usepackage{bm}
\usepackage{color}
\usepackage[utf8]{inputenc}
\usepackage{mathtools}
\usepackage[normalem]{ulem}

\theoremstyle{plain} %% This is the default

\definecolor{brickred}{rgb}{0.7, 0.25, 0.33}
\definecolor{applegreen}{rgb}{0.55, 0.71, 0.0}

\begin{document}

\title{Global topological  synchronization  on simplicial and cell complexes}

\author{Timoteo Carletti}
\affiliation{Department of Mathematics \& naXys, Namur Institute for Complex Systems, University of Namur, Rue Grafé 2, B5000 Namur, Belgium}
\author{Lorenzo Giambagli}
\affiliation{Department of Mathematics \& naXys, Namur Institute for Complex Systems, University of Namur, Rue Grafé 2, B5000 Namur, Belgium}
\affiliation{Department of Physics and Astronomy, University of Florence, INFN \& CSDC, Sesto Fiorentino, Italy}
\author{Ginestra Bianconi}
\affiliation{School of Mathematical Sciences, Queen Mary University of London, London, E1 4NS, United Kingdom}
\affiliation{The  Alan  Turing  Institute,  96  Euston  Road,  London,  NW1  2DB,  United  Kingdom}

\begin{abstract}
{Topological signals, i.e., dynamical variables defined on nodes, links, triangles, etc. of higher-order networks, are attracting increasing attention. However the investigation of their collective phenomena is only at its infancy. Here we combine topology and nonlinear dynamics to determine the conditions for global synchronization of topological signals {defined} on simplicial {or} cell complexes. On simplicial complexes we show that topological obstruction impedes odd dimensional signals to globally synchronize.  On the other hand, we show that cell complexes can overcome topological obstruction and in some structures, signals of any dimension can achieve global synchronization.}
\end{abstract}

\maketitle

Synchronization is a widespread phenomenon at the root of several biological rhythms or human made technological systems ~\cite{Pikovsky2001,arenasreview}. Synchronization refers to  the spontaneous ability of coupled oscillators to operate at unison and thus exhibit a coherent collective behavior. Global synchronization is the resulting phenomenon where all oscillators behave in the same way.

Traditionally synchronization has been studied when identical \cite{Pecora,Pecora_etal97} or non-identical oscillators \cite{Kuramoto,rodrigues2016kuramoto} are defined on the nodes of a network and are coupled by the network links.
However, to capture the function of many complex systems, e.g., {brain networks~\cite{giusti2016two,reimann2017cliques}, social networks~\cite{patania2017shape} and protein interaction networks~\cite{estradaJTB}},  it is important to go beyond pairwise interactions and consider  higher-order interactions \cite{battiston2021physics} between two or more nodes instead. {For instance collaborations typically require the cooperation of a team of more than two individuals, and to perform a function in the cell, proteins form complexes formed by several different types of proteins.}
Many-body interactions are captured by higher-order networks \cite{battiston2020networks,bianconi2021higher} (such as hypergraphs, simplicial and cell complexes) and are dramatically changing our understanding of the interplay between structure and dynamics of complex systems~\cite{bianconi2021higher,battiston2021physics,majhi2022dynamics}. {Note that higher-order networks by definition are constructed  by higher-order building blocks, but like networks they can have different structure~\cite{bianconi2021higher}. In particular a higher-order network can display a very regular (lattice-like) structure, a fractal \cite{Ziff} or a very random structure \cite{Owen}. Moreover higher-order networks can be also built from network motifs according to clique-complex construction, by identifying each clique with a simplex, and its generalizations~\cite{bianconi2021higher}.}

{Lately, synchronization of identical and non-identical oscillators defined on the nodes of  higher-order networks has been a field of intense research activity. Global synchronization of identical oscillators have been first formulated for special topologies (a  $p$-regular hypergraph) ~\cite{Krawiecki2014} and for a peculiar Laplace operator obtained from the hyper-adjacency matrix~\cite{MKJ2020} while recently a general and comprehensive theoretical framework has been proposed in~\cite{carlettifanellinicoletti2020} to study dynamical systems defined on hypergraph with heterogeneous hyperedges size distribution, the latter influencing also the Laplace matrix.}{ Partial synchronization of non-identical nodes oscillators has been investigated using a variation of the Kuramoto model leading to explosive transitions~\cite{skardal2019abrupt}.}

Simplicial and cell complexes also sustain {\em topological signals}~\cite{bianconi2021higher,PhysRevLett.124.218301,GRTB2021,sardellitti2021topological}, i.e., dynamical variables that can be defined not only on nodes, but also on links, triangles and higher-dimensional simplices or cells. Examples of real topological signals are edge signals such as biological transportation fluxes or traffic signals~\cite{sardellitti2021topological}, synaptic and brain edge signals~\cite{faskowitz2022edges}, or climate data such as currents in the ocean or speed of wind at different locations~\cite{schaub2021signal}. {In the framework of quantum systems, it has been shown that synchronization and quantum entanglement are strongly intertwined~\cite{MGGHZ2013,WWBT2017}; observation that can be relevant in quantum computations involving anyons~\cite{KITAEV20032} or bosons~\cite{SemeghiniEtAl2021} where signals defined on links of some array, should be protected from noise.} {Finally, topological signals} are  attracting increasing attention in signal processing and machine learning~\cite{sardellitti2021topological,schaub2021signal,bodnar2021weisfeiler,ebli2020simplicial,sardellitti2021topological,calmon2023dirac}. However the study of their collective phenomena its only at its infancy~\cite{PhysRevLett.124.218301,GRTB2021,torres2020simplicial,ziegler2022balanced,arnaudon2022connecting,calmon2021topological}.

{Recently the formulation of  higher-order topological Kuramoto model~\cite{PhysRevLett.124.218301,GRTB2021,calmon2021topological} has demonstrated that topological signals of any dimension can synchronize leading {to either continuous or to}  explosive synchronization transitions. These results concern  partial synchronization while an important question is whether global synchronization of topological signals can ever be achieved.}

{The aim of this work is to determine the topological and dynamical conditions under which global topological synchronization of identical topological oscillators can be observed.} Relying on the use of higher-order Laplacian matrices \cite{bianconi2021topological,GradyPolimeni2010,lim2020hodge,horak2013spectra}  and the extension of the Master Stability Function (MSF) to simplicial and cell complexes dynamics, we are able to tackle this problem  emphasizing the difference existing among the two frameworks.

Anticipating on {our results} we can state that on  {simplicial complex we observe topological obstruction}:  given a simplicial complex of dimension $K$, if the topological signal is defined on an odd-dimensional simplex of dimension $k<K$ then, global synchronization is not possible. On the other hand if the simplex has an even dimension, then we can have global synchronization provided the simplex is {\em balanced} (see hereafter) and the model parameters allow for {it}. {Interestingly we show that cell complexes can overcome topological obstruction and some topologies can sustain global synchronization of signals of any dimension. }

\noindent{\em Definition of simplicial and cell complexes.}-- Simplicial and cell complexes are generalised network structures that beside nodes and links do also contain triangles, polygons, tetrahedra, hypercubes, orthoplexes and so on. Given a non-negative integer $k$, a {\em  $k$-simplex}, $\sigma^{k}$, is {a set} of $k+1$ different {nodes,  $\sigma^k=[v_0,v_1,\ldots, v_k]$.} A $(k-1)$-face, $\sigma^{k-1}$, of a $k$-simplex is a $(k-1)$-simplex obtained by removing one vertex from $\sigma^{k}$. {Therefore a node is a $0$-simplex, a link is a $1$-simplex,  a $2$-simplex is a triangle and so on. } A simplicial complex $\mathcal{X}$ of dimension $K$ is a finite collection of simplices of dimensions $k\leq K$, closed under the inclusion of faces, namely if $\sigma \in \mathcal{X}$, then also all the faces of $\sigma$ should belong to $\mathcal{X}$. Simplicial complexes are, thus a natural generalisation of networks, {which are recovered for $k=1$.} {In algebraic topology a simplex $\sigma^k$ is assigned an {\em  orientation}, typically induced by the node labels}. A coherent orientation of the face with the one of the simplex, will be denoted by $\sigma^{k-1}\sim \sigma^{k}$, otherwise we will write $\sigma^{k-1}\not\sim \sigma^{k}$. 

A more general structure retaining the algebraic richness of simplicial complexes, is given by the {\em cell complexes}~\cite{mulder2018network,Steinitz1908,Klette2000,Hatcher2005,GradyPolimeni2010}. The latter differs from simplicial complexes because they are not just built by simplices but instead they are obtained by gluing cells, (i.e., regular polytopes) along their faces. In particular $0$-cells are nodes, $1$-cells links, while $2$-cells are generic polygons,{and $3$-cells are the Platonic polytopes}. 

  {In algebraic topology the boundary operator is defined on chains \cite{bianconi2021higher,GradyPolimeni2010}, linear combinations of oriented simplices of the simplicial complex. The $k$ boundary operator retains the information about the faces of a $k$-simplex $\sigma^k$ and their relative orientation.} 
  %e.g., two simplices of order $k$, $\sigma_1^{k}$ and $\sigma_2^{k}$ are lower adjacent if they have a common face of order $k-1$; they are upper adjacent if both are faces of a simplex of dimension $k+1$. 
  The boundary operator is  encoded into the {\em incidence matrices}, $\mathbf{B}_k$ whose elements are given for all $k=1,\dots,K$ by $B_k(i,j)=1$ if $\sigma_i^{k-1}\sim \sigma_j^{k}$, $-1$ if $\sigma_i^{k-1}\not\sim \sigma_j^{k}$ and $0$ otherwise~\footnote{Let us observe that the $(i,j)$ entry of the matrix $\mathbf{B}_k$ has been denoted by $B_k(i,j)$; a similar assumption has been used to denote the entries of vectors.},
%\begin{equation}
%\label{eq:Bk}
%B_k(i,j)=
%\begin{cases}
% 1 & \text{ if } \sigma_i^{k-1}\sim \sigma_j^{k}\\
% -1 & \text{ if } \sigma_i^{k-1}\not\sim \sigma_j^{k}\\
%  0 & \text{ otherwise}
%\end{cases}\, k=1,\dots,K,
%\end{equation}
 where $K$ is the dimension of the simplicial complex, namely the size of the largest simplex. One key property of the boundary operator is that the {\em boundary of the boundary is null}, hence $\mathbf{B}_k\mathbf{B}_{k+1}=0$. To be coherent an oriented simplicial complex needs to satisfy this condition for every $k$ with $0\leq k\leq K$. Let us conclude this section by noticing that the boundary operators {for cell complexes have the same definition} \cite{sardellitti2021topological}.

The higher-order Laplacians \cite{bianconi2021higher,torres2020simplicial,GradyPolimeni2010,lim2020hodge,horak2013spectra} allow to {define diffusion among $k$-simplices and capture the topology of the higher-order network}; they are defined in terms of the incidence matrix as:
\begin{equation}
\label{eq:L}
\mathbf{L}_k = \mathbf{B}_k^\top\mathbf{B}_{k}+\mathbf{B}_{k+1}\mathbf{B}_{k+1}^\top \, , \quad k=1,\dots, K-1\, .
\end{equation}
For $k=0$ and $k=K$ we have instead $\mathbf{L}_0 = \mathbf{B}_1\mathbf{B}_{1}^\top$ and $\mathbf{L}_K = \mathbf{B}_K^\top\mathbf{B}_{K}$. If we denote by $N_k$, $k=0,\dots, K$, the number of $k$-simplices, then it follows that $\mathbf{B}_{k}$ is a $N_{k-1}\times N_k$ matrix, while the size of $\mathbf{L}_{k}$ is $N_{k}\times N_k$. The matrix $\mathbf{L}_{0}$  coincides with the {graph Laplacian}. 
{The higher-order Laplacian $\mathbf{L}_k$ is a semidefinite operator and one of its most celebrated properties is that the dimension of its kernel is given by the $k$-Betti number, i.e., $\mathrm{dim}(\ker\mathbf{L}_k)=\beta_k$, where $\beta_k$ indicates the number of  $k$-dimensional cavity in the simplicial complex. Moreover there is a basis of the kernel $\mathbf {L}_k$ which is  formed by eigenvectors localized on each of the $k$-dimensional cavities of the simplicial complex.}

\noindent{\em Simplicial and cell-complex dynamics.}-- Let us now consider a topological $k$-dimensional signal {encoded in a $k$ dimensional cochain $\mathbf{x}:C_k\to \mathbb{R}^d$ which assigns to every chain $C_k$ (linear combination of $k$-simplices) values on $\mathbb{R}^d$. The $k$-topological signal has elements $\mathbf{x}_i=\mathbf{x}(\sigma_i^k)=(x_i^1,\dots,x_i^d)$ defined on the $i$-th  oriented $k$-simplex, $\sigma_i^k$ (see SM). According to the properties of the $k$-cochains \cite{GradyPolimeni2010,lim2020hodge} we have ${\bf x}(-\sigma_i^k)=-{\bf x}(\sigma_i^k)$, being the discrete analogous of differential forms on manifolds.} For instance, for $k=1$ {and $d=1$}, $x_i={x}(\sigma_i^k)$  indicates a flux defined on the link $i$ that is positive if going  in the same  direction of the positive orientation of the link and negative otherwise, i.e., $x(-\sigma_i^k)=-x_i$.  Let us assume the value of the topological signal on every simplex $i$ follows the same dynamics and  evolves according to $\dot{\mathbf{x}}_i=\mathbf{f}(\mathbf{x}_i)$, {for some odd nonlinear function $\mathbf{f}:\mathbb{R}^d\rightarrow \mathbb{R}^d$.} 
Assume now the $k$-simplex to belong to a $K$-simplicial complex, $K\geq k$, and assume the existence of a diffusive--like nonlinear interaction among adjacent simplices of the same dimension:
\begin{equation}
\label{eq:coupled}
\frac{d\mathbf{x}_i}{dt} = \mathbf{f}(\mathbf{x}_i)-\sum_{j=1}^{N_k} L_k(i,j) \mathbf{h}(\mathbf{x}_j)\quad\forall i=1,\dots, N_k\, ,
\end{equation}
where {$\mathbf{h}:\mathbb{R}^d\rightarrow \mathbb{R}^d$ is some odd nonlinear coupling function.}  
This equation generalizes the dynamics of identical oscillators anchored to each node~\cite{Pecora} to the scenario in which identical oscillators are defined on higher-dimensional simplices or cells. 
{Please note that requiring odd functions $\mathbf{f}(\mathbf{x}_i)$ and $\mathbf{h}(\mathbf{x}_i)$ is necessary for higher-order topological signals with $k>0$ in order to ensure invariance under change of orientation of each simplex $i$ (see SM).} For
node dynamics ($k=0$) the existence of a global synchronized state is automatically determined by the properties of the graph Laplacian whose kernel is spanned by $\mathbf{u}=(1,\ldots, 1)^{\top}$, indeed we have $\beta_0=1$, one connected component. Its stability is instead determined by the celebrated MSF~\cite{Pecora,Pecora_etal97}.

{Given the growing interest in topological signals, a key question is how these classic results of nonlinear dynamics on networks extend to nonlinear dynamics of topological signals on simplicial complexes. Anticipating our results, we will show that topology and  combinatorics of the higher-order Laplacian will not always ensure existence of a globally synchronized state, and moreover since the dimension of the kernel of $\mathbf{L}_k$ can be bigger than one, also the MSF will differ from the network case.}

Let us then fix a reference stable solution $\mathbf{s}(t)$ of the uncoupled system, $\dot{\mathbf{x}}_i=\mathbf{f}(\mathbf{x}_i)$. We are interested {in determining the conditions under which the state having each simplex $i$ either in the state  $\mathbf{x}_i=\mathbf{s}(t)$ or in  $\mathbf{x}_i=-\mathbf{s}(t)$ is also a stable solution of the coupled system~\eqref{eq:coupled}. Namely the latter exhibits a {\em global synchronous} behavior in which all simplices display the same dynamics of the isolated simplices when we account for differences of sign, determined by their orientation (see SM).}

Let us now introduce the vector $\mathbf{v}=(v_1,\dots,v_{N_k})^\top\in \{-1,1\}^{N_k}$, 
such that the globally synchronized state is given by $\mathbf{x}_i=v_i\mathbf{s}(t)$. A {\em necessary condition} {to observe global synchronization,} is that $\sum_j L_k(i,j)\mathbf{v}_j=0$  {(see SM).} Let us recall that $\mathrm{ker}\mathbf{L}_k=\mathrm{ker}\mathbf{B}_k\cap \mathrm{ker}\mathbf{B}_{k+1}^\top$, thus the latter condition ultimately returns to require $\mathbf{B}_k\mathbf{v}=0$ (condition \textbf{i}) and $\mathbf{v}^{\top}\mathbf{B}_{k+1}=0$ (condition \textbf{ii}) (see SM).
\begin{figure*}[ht]
\centering
\includegraphics[scale=0.21]{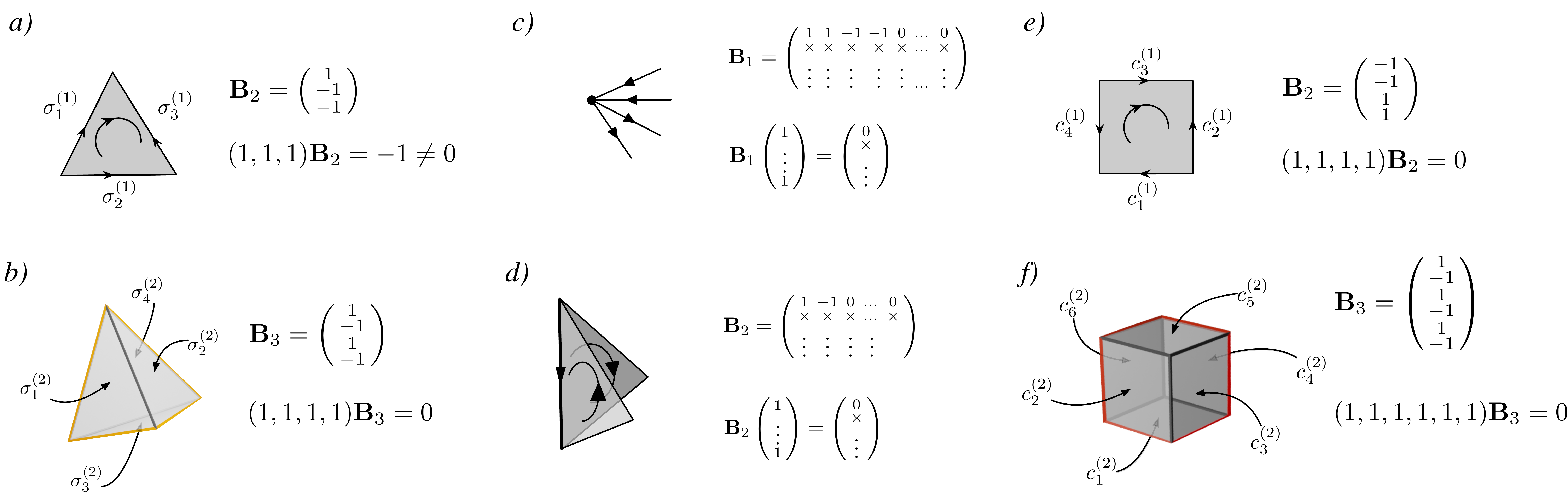}
%\vspace{-1.5cm}
\caption{{Schematic description of conditions \textbf{i)} (panels a,b,e,f) and \textbf{ii)} (panels c and d) for topological signals defined on $1$-dimensional cells (links, panels in top row) and $2$-dimensional cells (triangles or squares, panels in bottom row) in which we assume that there is an orientation such that $\mathbf{w}=\mathbf{u}=(1,1,\ldots)^{\top}$. In the case of simplicial complexes (panels a, d) condition \textbf{i)} cannot be satisfied for signals defined on $1$-dimensional simplices while in the case of   cell complexes (panels e, f) condition \textbf{i)} it can be satisfied. Condition \textbf{ii)} can be satisfied on simplicial and cell complexes as long as the simplices are balanced (see panels c, d) for the simplicial complex case).}}
\label{fig:condupdown}
\end{figure*}

The first condition has a striking consequence. If $k$ is an odd number, because any $(k+1)$-simplex contains an odd number of $k$-faces, then  condition \textbf{i)} cannot be ever satisfied. On the contrary if $k$ is even, then any $(k+1)$-simplex contains an even number of $k$-faces  condition \textbf{i)} can be realised (see Fig.~\ref{fig:condupdown}a-b). On the other hand condition \textbf{ii)} can be satisfied by imposing a suitable condition of the $(k-1)$-faces of the $k$-simplex, which we call {\em balanced condition}. In particular if $\mathbf{v}=\mathbf{u}=(1,\ldots,1)^{\top}$ this condition can be satisfied by requiring every $(k-1)$-face to be adjacent to an even number of $k$-simplices and to be coherently oriented with half of them (see Fig.~\ref{fig:condupdown}c-d).

Therefore for even values of $k$, global synchronization can be achieved while if $k$ is odd we observe, as long as $\mathbf{v}^{\top}{\bf B}_{k+1}\neq {\bf 0}$, a {\em topological obstruction} to the onset of global synchronization. Interestingly for $K$-dimensional signals  having ${\bf B}_{K+1}={\bf 0}$, only the balanced condition remains (i.e., condition \textbf{ii)}) which is automatically satisfied for the  vector $\mathbf{v}=\mathbf{u}$ if the simplicial complex is a closed manifold without boundary. Hence $K$-dimensional topological signals defined  on closed $K$ dimensional manifolds can always achieve global synchronization for arbitrary value of $K$.

{A similar derivation can be generalized and extended to topological signals defined on the $k$-dimensional cells of cell complexes. In particular the conditions to achieve global synchronization on a cell complex are unchanged and given again by condition ${\bf i)}$ and ${\bf ii)}$. However the combinatorics of cell complexes is different from the one of simplicial complexes. Take for instance a cell complex whose network skeleton is formed by a $d$-dimensional square lattice with periodic boundary conditions, i.e., a regular tessellation of $d$-dimensional torus. Then every cell of dimension $k+1>0$ has a even number of $k$-dimensional faces therefore condition ${\bf i)}$ can be satisfied also if $k$ is odd (see Fig.~\ref{fig:condupdown}e-f).  This implies that on cell complexes we can  overcome topological obstruction.}
Until now we have focused on the combinatorial implication of the conditions $\mathbf{i)}$ and $\mathbf{ii)}$. However these  conditions have also topological consequences. In fact, since on manifolds the eigenvectors of the kernel of the Hodge Laplacian $\mathbf{L}_k$ are localized on the $k$-dimensional holes, manifolds that will display an  eigenvector with the properties of the above defined $\mathbf{v}$, are characterized by holes spanning the whole structure as for instance $(k+1)$-dimensional hyperspheres or $d$-dimensional tori with $d>k$.

\noindent{\em Master Stability Equation for Topological Signals.}--
Let us now assume the reference solution $\mathbf{s}(t)$ to be also a solution of the coupled system~\eqref{eq:coupled}, then by introducing the distance from the reference orbit, $\delta\mathbf{x}_i=\mathbf{x}_i-\mathbf{s}(t)$, we can derive its time evolution by linearizing Eq.~\eqref{eq:coupled}:
\begin{equation}
\label{eq:coupledlin}
\frac{d\delta\mathbf{x}_i}{dt} = \mathbf{J}_\mathbf{f}(\mathbf{s})\delta\mathbf{x}_i-\sum_{j=1}^{N_k} L_k(i,j) \mathbf{J}_\mathbf{h}(\mathbf{s})\delta\mathbf{x}_j \quad\forall i=1,\dots, N_k\, ,\nonumber
\end{equation}
being $\mathbf{J}_\mathbf{f}(\mathbf{s})$ (resp. $\mathbf{J}_\mathbf{h}(\mathbf{s})$) the Jacobian of the function $\mathbf{f}$ (resp. $\mathbf{h}$) evaluated on the reference solution.

The matrix $\mathbf{L}_k$ being symmetric, it admits an orthonormal basis, $\pmb{\phi}_k^{(\alpha)}$, associated to eigenvalues $\Lambda_k^{(\alpha)}$, $\alpha=1,\dots, N_k$, namely $\mathbf{L}_k\pmb{\phi}_k^{(\alpha)}=\Lambda_k^{(\alpha)}\pmb{\phi}_k^{(\alpha)}$. In particular, {since we work under the assumption that the simplicial complex is  balanced}, $\pmb\phi_k^{(1)}\sim (1,\dots,1)^\top\in\mathbb{R}^{N_k}$, $\Lambda_k^{(\alpha)}=0$ for {$1\leq \alpha\leq \beta_k$ and $\Lambda_k^{(\alpha)}>0$ for all $\alpha >\beta_k$.}

Let us decompose the deviation vectors $\delta\mathbf{x}_i$ onto this eigenbasis: $\delta\mathbf{x}_i=\sum_\alpha \delta\mathbf{x}^{(\alpha)}\pmb\phi_{k}^{(\alpha)}(i)$. Then linearizing the dynamical equation, we obtain
\begin{equation}
\label{eq:coupledlinspect}
\frac{d\delta\mathbf{x}^{(\alpha)}}{dt} = \left[\mathbf{J}_\mathbf{f}(\mathbf{s})-\Lambda_k^{(\alpha)}\mathbf{J}_\mathbf{h}(\mathbf{s})\right]\delta\mathbf{x}^{(\alpha)}\quad \forall \alpha=1,\dots,N_k\, .\nonumber
\end{equation}
{Perturbations aligned with the kernel do not change the stability of the uncoupled system, therefore only the perturbations orthogonal to the kernel can modify the stability of the reference solution.} 
This is the MSF in the framework of simplicial synchronization of topological signals. It is a linear, in general non autonomous, ODE parametrized by the eigenvalues $\Lambda_k^{(\alpha)}$, allowing to infer the stability character of the reference solution by looking at its spectrum.

\begin{figure}[ht]
\centering
\includegraphics[scale=0.16]{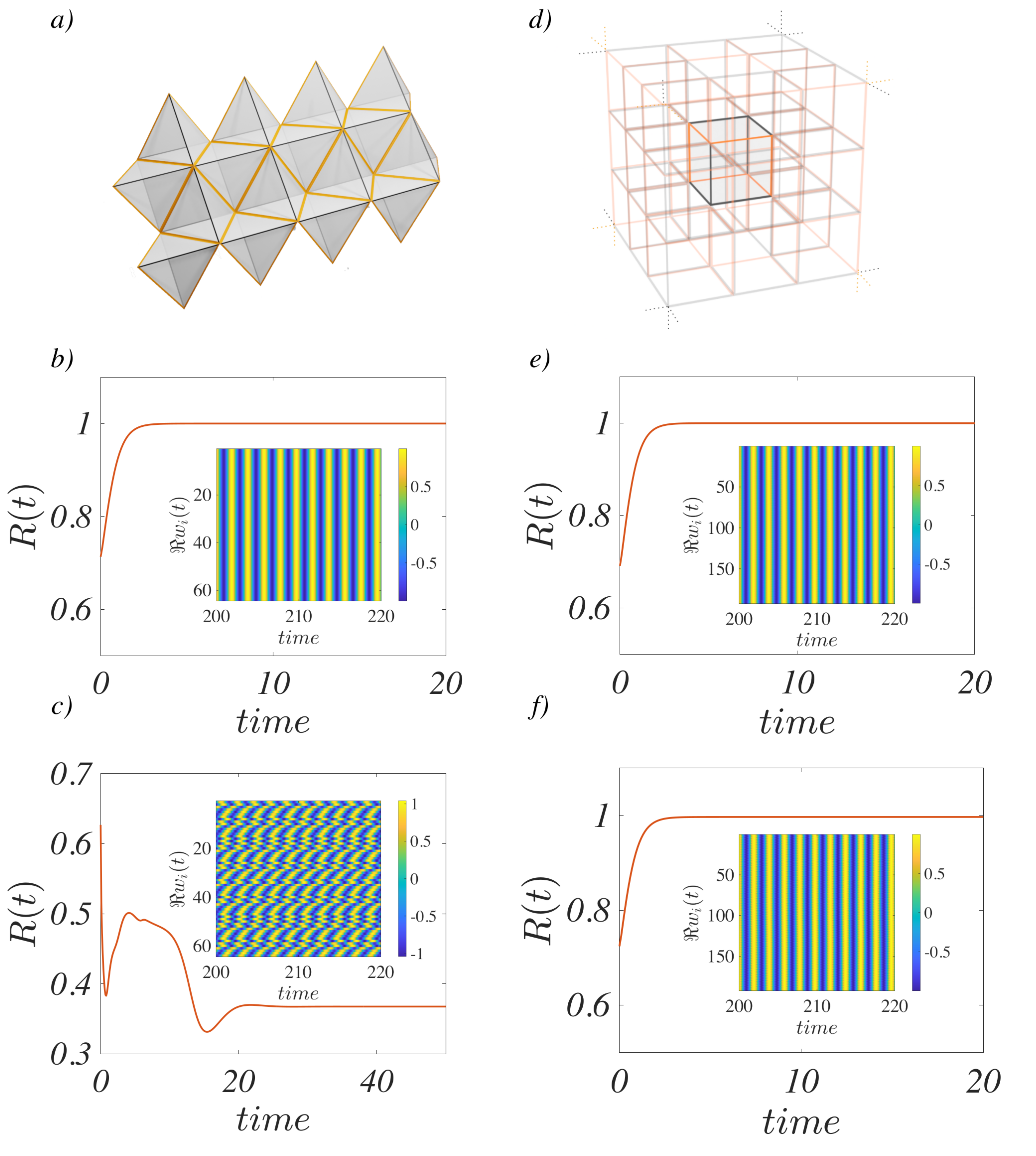}
\caption{{The Kuramoto order parameter $R$ is plotted versus time $t$ for the Stuart-Landau model of topological oscillators of the balanced simplicial and cell complexes represented in panels a) and d) respectively.
Panels b) and c) refer respectively to the order parameter of triangles and links of the simplicial complex in panel a). Panels e) and f) refer respectively to the order parameter of the squares and links of the cell complex in panel d). The insets display the dynamical time series of the topological signals.
It is clear that while on the links of the simplicial complex, the oscillators do not globally synchronize, the ones of the cell complex do support synchronization.
The model parameters are $\sigma = 1.0+4.3 i$, $\beta = 1.0+1.1 i$, $\mu = 1.0-0.5i$ and $m = 3$ ensuring the negativity of the dispersion relation (see SM).}}
\label{fig:SLsmplx}
\end{figure}
\noindent{\em Simplicial Stuart-Landau model.}-- {As an application of the general theory above introduced, let us consider the Stuart-Landau (SL) model~\cite{vanharten,aranson,garcamorales} defined for topological signals of dimension $k$ and $d=2$. {For $k=0$ the model describes a nonlinear oscillator anchored at each node, while for $k=1$ it can describe an oscillating flux associated to an edge linking two nodes.} More precisely let us define {$w_j=x_j^1+\mathrm{i}x_j^2$ and let us consider} the ``local reaction'' function {$f({\bf x})=\tilde{f}(w)=\sigma w -\beta w|w|^2$,} where $\sigma=\sigma_\Re+i\sigma_\Im$ and $\beta=\beta_\Re+i\beta_\Im$ are complex control parameters. 
We can prove that the uncoupled dynamics $\dot{w}_j=f(w_j)$ in each simplex $j$ admits a limit cycle solution $\hat{z}(t)=\sqrt{\sigma_\Re/\beta_\Re}e^{i\omega t}$, where $\omega=\sigma_\Im-\beta_\Im \sigma_\Re/\beta_\Re$, that is stable provided $\sigma_\Re>0$ and $\beta_\Re>0$, conditions that we hereby assume. 
We now consider the coupled dynamics  Eq.~\eqref{eq:coupled} with  nonlinear coupling function {$h({\bf x})=\tilde{h}(w)=\mu w |w|^{m-1}$}, where $m$ is a positive integer and  
 $\mu=\mu_\Re+i\mu_\Im$ a complex parameter that sets the coupling strength~\cite{CarlettiFanelli2022}.}
We study the  stability of the reference limit cycle solution $\hat{z}(t)$  (see SM) and we prove that the system can globally synchronize, i.e., the dispersion relation is negative,  only if the model parameters do satisfy
$\mu_\Re +\mu_\Im {\beta_\Im}/{\beta_\Re}>0\, ,$
and the simplicial complex is such that $\mathbf{u}\in\ker \mathbf{L}_k$. To measure global synchronization we compute the (generalised) Kuramoto order parameter 
$R(t)=\Big\lvert\sum_j \rho_j(t)e^{i \theta_j(t)}\Big\rvert\ /N_k$, 
where $w_j(t)=\rho_j (t)e^{i\theta_j(t)}$ is the polar form of the complex signal. Then $R(t)\rightarrow 1$ testifies the existence of phase and amplitude synchronization. Results shown in  Fig.~\ref{fig:SLsmplx} provide numerical evidence of our theoretical predictions. In Fig.~\ref{fig:SLsmplx}a-c we show the results obtained by studying the SL model defined on top of {a designed balanced $3$-simplicial complex} (see SM). The model parameters have been set to values allowing for a negative dispersion relation (see SM) and indeed once the complex amplitudes are defined on $2$-faces, i.e., triangles, the system globally synchronizes (see Fig.~\ref{fig:SLsmplx}b). On the other hand, once the SL oscillators are defined on links the system cannot globally synchronize (see Fig.~\ref{fig:SLsmplx}c). {In Fig.~\ref{fig:SLsmplx}d-f we provide an example of a cell complex which overcomes topological obstruction: a $3D$ square lattice with periodic boundary conditions. Such cell complex is made of nodes, link, squares and cubes (see SM). In this case  case global synchronization can be achieved for signals of every dimension (see Fig.~\ref{fig:SLsmplx}e-f for global synchronization of links and squares).}

{In conclusion in this work we have studied global synchronization of identical topological oscillators on simplicial or cell complexes. We have found that global synchronization of topological signals cannot be observed on arbitrary simplicial or cell complexes but that only some special higher-order network topologies can sustain such a dynamical state. This is in stark contrast with the corresponding scenario in networks where global synchronization can be observed in every network topology given proper dynamical conditions.
By combining topology, and in particular the spectral properties of higher-order Laplacians, to nonlinear dynamics techniques such as the MSF,
we have identified the topological and dynamical conditions under which identical topological oscillators can achieve global synchronization on simplicial or cell complexes.
We have proved that global synchronization of odd dimensional topological signals is obstructed in simplicial complexes. This topological obstruction implies that on a $K$-dimensional simplicial complex we can never observe global synchronization of  odd dimensional topological signals of dimension $d<K$. However, such obstruction is not present in cell complexes. In particular we show evidence that in specific topologies such as the $d$-dimensional square lattice with periodic boundary conditions global synchronization of topological signal of any dimension can be observed.}

{These  results  significantly enrich our  understanding of the relation between higher-order network topology and dynamics revealing collective phenomena of topological signals. Our study is relevant, for its inherent simplicity, to a wide spectrum of applications  (neuroscience/biology/social sciences) where many-body interactions involve higher-order interacting units. Therefore we hope that this Letter will stimulate further research work in physics and beyond.}

G.B. acknowledges funding from the Alan Turing Institute  and the Royal Society (IEC\textbackslash NSFC\textbackslash191147).

\bibliography{bib_RTP}

\newpage
\onecolumngrid
\appendix

\renewcommand\theequation{{S-\arabic{equation}}}
\renewcommand\thetable{{S-\Roman{table}}}
\renewcommand\thefigure{{S-\arabic{figure}}}
\setcounter{equation}{0}
\setcounter{figure}{0}
\setcounter{section}{0}
\begin{center}
\Large{\bf SUPPLEMENTAL MATERIAL}
\end{center}
\section{{Extension of the Master Stability Function Approach for  dynamics of  topological signals}}
\label{sec:sysdyncmplx}

In this section we provide the mathematical background for the extension of the Master Stability Function (MSF) defined on networks to the MSF for topological signals of a simplicial or cell complex.

Let us start by defining a $k$-dimensional chain $c\in \mathcal{C}_k$ as a linear combination of oriented simplices (or cells) $\sigma^k_i$, i.e.
\begin{equation}
    c=\sum_i c_i \sigma^k_i
\end{equation}
with coefficients $c_i\in \mathbb{R}$.

A $k$-dimensional topological signal $x$ is a $k$-dimensional cochain $x:\mathcal{C}_k\to \mathbb{R}^d$, i.e., a linear function that associates to every $k$-chain of the simplicial complex a value in $\mathbb{R}^d$.

Since the  cochain indicates a linear function we have 
\begin{equation}
    x(c)=\sum_i c_ix(\sigma^k_i).
    \label{uno}
\end{equation}
Note that thanks to the linearity of the cohain $x$ we always have $x(\sigma^k_i)=-x(-\sigma^k_i)$.
From Eq.(\ref{uno}) it follows that a cochain $x$ is uniquely determined by the vector $\mathbf{x}$ of elements $\mathbf{x}_i=x(\sigma^k_i)$.

{Let us consider a cochain $\mathbf{x}$ and a function $\mathbf{F}$ that associated to $\mathbf{x}$ a new cochain denoted by $\mathbf{F}(\mathbf{x})$. Based on the above, the latter is solely characterized by its values on $\sigma_i^k$, namely its ``components'', $\mathbf{F}(\mathbf{x})=(\mathbf{f}_1(\mathbf{x}),\dots,\mathbf{f}_{N_k}(\mathbf{x}))$. Let us finally assume each component to act componenent-wise, namely $\mathbf{f}_i(\mathbf{x})=\mathbf{f}_i(\mathbf{x}_i)$ for all $i$. Being $\mathbf{x}$ and $\mathbf{F}(\mathbf{x})$ cochains they should be invariant with respect to the change of orientation of $\sigma_i^k$, namely}
\begin{equation*}
{\mathbf{f}_i(\mathbf{x}_i)=-\mathbf{f}_i(-\mathbf{x}_i)\, .}
\end{equation*}
 
We now want  to study the evolution of the topological signals by   considering dynamical equations for the vector $\mathbf{x}$.

The uncoupled dynamics of topological signals is determined by 
\begin{equation}
    \frac{d\mathbf{x}_i}{dt}=\mathbf{f}({\bf x}_i),
    \label{due}
\end{equation}
where $\mathbf{f}:\mathbb{R}^d\to \mathbb{R}^d$ is a nonlinear function which should preserve the invariance of the dynamics under  a change of the  orientations of the simplices.
Requiring the invariance of this equation under change of orientation of the simplex $i$, imposes that  the function $\mathbf{f}(\mathbf{x})$ must be odd.
Indeed if the orientation of the simplex $\sigma^k_i$ is reversed we have $\mathbf{x}_i\to-\mathbf{x}_i$ and the dynamical Eq.(\ref{due}) becomes
\begin{equation}
    -\frac{d\mathbf{x}_i}{dt}=\mathbf{f}(-{\bf x}_i).
\end{equation}
{This is indeed the above presented case where $\mathbf{f}_i=\mathbf{f}$ for all $i$.} Therefore to  ensure that the dynamical equation (\ref{due}) obeys the necessary invariance conditions for describing the dynamics of topological signals of dimension $k>0$,
 we need to require $\mathbf{f}(-{\bf x}_i)=-\mathbf{f}({\bf x}_i)$, hence that $\mathbf{f}(\mathbf{x}_i)$ is an odd function.  Note that requiring odd nonlinear functions $\mathbf{f}(\mathbf{x}_i)$ is a necessary condition  only for higher-order topological signals  of dimension $k>0$. Indeed since nodes do not have an orientation this requirement is not necessary for treating topological signals defined on nodes. Therefore this is an important difference with the dynamics defined exclusively on nodes.

Let us now consider a stable solution $\mathbf{s}(t)$ of Eq.(\ref{due}). Due to the parity of $\mathbf{f}(\mathbf{x})$ we are automatically guaranteed that also $-\mathbf{s}(t)$ is a stable  solution of Eq.(\ref{due}).

We define as {\em global synchronization} of topological signals a dynamical configuration in which $\mathbf{x_i}=\mathbf{s}(t)$ or $\mathbf{x_i}=-\mathbf{s}(t)$ for every $k$-dimensional simplex $i$ of the simplicial or cell complex. This definition again differs from the one of global synchronization defined for node signals and guarantees the invariance under the orientation of the simplices. To give an intuition of this result, take the case of link signals indicating fluxes: saying that a flux is $\phi$ on the link $[i,j]$  oriented from $i$ to $j$ is the same as saying that the flux is $-\phi$ on the link $[i,j]$ oriented from $j$ to $i$. Therefore global synchronization is observed where the dynamics of each simplex is the same, when factoring out the sign due to the choice of orientation.

We now turn our attention to the case of interacting topological signals coupled by the Hodge Laplacian operator whose dynamics is dictated by Eq.(3) in the main text that we rewrite here for convenience 
\begin{equation}
    \frac{d\mathbf{x}_i}{dt} = \mathbf{f}(\mathbf{x}_i)-\sum_{j=1}^{N_k} L_k(i,j) \mathbf{h}(\mathbf{x}_j)\quad\forall i=1,\dots, N_k\, ,
    \label{eq:top_oscillators}
\end{equation}
where $\mathbf{h}:\mathbb{R}^d\to \mathbb{R}^d$ is a nonlinear function which should preserve the invariance under choice of the orientation of the simplices (or cells) and where $\mathbf{L}_k$ is the $k$-dimensional Hodge Laplacian.
Taking into account that the latter has elements
\begin{equation}
    L_k(i,j)=\left\{\begin{array}{ll}
d(\sigma^k_i)+(p+1), & i=j. \\
1, & i\ne j, \sigma^k_i \not\frown \sigma^k_j, \sigma^k_i \smile \sigma^k_j, \sigma^k_i \sim \sigma^k_j. \\
-1, & i\ne j, \sigma^k_i \not\frown \sigma^k_j, \sigma^k_i \smile \sigma^k_j, \sigma^k_i \nsim \sigma^k_j. \\
0, &  \text{otherwise}
\end{array} \right.,
\end{equation}
where $\smile$ and $\frown$ indicate  lower  and upper incident neighbor simplices respectively,
it can be easily derived that in order to preserve the mentioned invariance also $\mathbf h(\mathbf{x})$ must be an odd function.
Note that this is another example of a  peculiar property of higher-order topological signals with $k>0$ which is not present for the description of dynamics of identical oscillators placed on the  nodes (node  signals: case $k=0$).

We now address the question of where global synchronization of higher-order topological signals can be observed.
From Eq.~\eqref{eq:top_oscillators} it is clear that a necessary condition to observe global synchronization is that the kernel of the Hodge Laplacian $\mathbf{L}_k$ admits an eigenvector $\mathbf{v}$ whose elements are in $\{-1,1\}$, i.e., $|v_i|=1$, for all $i$.
Note that since the degeneracy of the zero eigenvalue of $\mathbf{L}_k$ is given by the $k$-th Betti number, it follows that the Hodge Laplacian $\mathbf{L}_k$, when $k>0$, can admit more than one eigenvectors of this form.
However we are not guaranteed that an arbitrary simplicial or cell complex will have at least one such eigenvectors.
In order to give a concrete example of such topologies we mention that   a $d$-dimensional regular cell complex which tessellates a torus  has $\text{Bin}(d,k)$ such eigenvectors while any simplicial or cell complex without any $k$-dimensional hole will have $k$-th Betti number $\beta_k=0$ and hence any such eigenvector.
This is a property that is  in strike contrast with what happens for $k=0$ where any connected network has one and only one constant eigenvector $(1,\ldots, 1)^{\top}$ in its kernel.

Let us consider a simplicial or cell complex which has an eigenvector $\mathbf{v}$ as above, therefore as stated in the main text, a necessary condition to have global synchronization is that the kernel of the Hodge Laplacian $\mathbf{L}_k$ admits the eigenvector $\mathbf{v}$.
This necessary condition is met under the combinatorial conditions \textbf{i}) and \textbf{ii}) discussed in the main text and for more detail in the next section of the SM. 
Note that these combinatorial constraints have also topological consequences as discussed in the main text, indeed they imply that global synchronization is allowed on topologies  having holes that span 
the entire structure as hypersphere and tori.

The Master Stability Equation (MSE) for topological signals is discussed in the main text. We observe that in this case the major difference with the MSE for node signals ($k=0$) is that the perturbations can also lie on the kernel since the $k$-th Betti number $\beta_k$ of the simplicial complex can be eventually also bigger than one. These perturbations however do not change the stability of the dynamics, since the non-interacting solutions is assumed to be stable.

\section{About the topological obstruction}
\label{sec:topobs}

Let us consider a $k$ dimensional topological signal whose time evolution is given by Eq.~\eqref{eq:coupled} (main text), we have shown that a necessary condition for the existence of the global synchronization manifold is that the vector $\mathbf{v}$ of elements $|v_i|=1$ lies in $\ker \mathbf{L}_k$. Because $\mathrm{ker}\mathbf{L}_k=\mathrm{ker}\mathbf{B}_k\cap \mathrm{ker}\mathbf{B}_{k+1}^\top$, such condition is equivalent to require $\mathbf{v}^\top\mathbf{B}_{k+1}=0$ (condition \textbf{i}) and $\mathbf{B}_k\mathbf{v}=0$ (condition \textbf{ii}).

Condition \textbf{i}) imposes a constraint on the dimension of the simplex. It cannot be verified on a $k$-simplex being $k$ an odd number smaller than the simplex dimension. Here we consider a pedagogical example with $k=1$, i.e., the topological signal is defined on a link. Because any $2$-simplex, i.e., a triangle, contains three links, whichever the orientation they have and for any choice of the vector $\mathbf{v}$, then $\mathbf{v}^{\top}\mathbf{B}_2$ cannot vanish, being a sum of an odd number of ``$+1$'' and ``$-1$'', thus condition \textbf{i}) cannot be satisfied see Fig.~\ref{fig:condupdown}a). On the other hand if $k$ is even we can build a suitable $k$-simplex verifying condition \textbf{i}). For simplicity we propose an example with $k=2$, i.e., a triangle. Then any tetrahedron must contain four triangles and we can always give them a suitable orientation, two by two coherent each other, in such a way the sum of the column of $\mathbf{B}_3$ is zero see Fig.~\ref{fig:condupdown}b).

On the other hand condition \textbf{ii}) can be satisfied by imposing the {\em balanced condition} on the $(k-1)$-faces of the $k$-simplex. Let us consider for instance the case in which $\mathbf{v}=\mathbf{u}=(1,\ldots, 1)^{\top}$. In Fig.~\ref{fig:condupdown}c we consider again, as an example, the case $k=1$, hence to satisfy condition \textbf{ii}) is enough to impose that each $(k-1)$-simplex, i.e., a node in this case, is incident with an even number of $1$-simplex (four in the figure), half of them coherently oriented, i.e., entering in the node, and half of them non coherently oriented, i.e., exiting from the node. The validity of this condition does not depend on the dimension of the simplex, indeed in panel d) we consider the case of $k=2$, i.e. a triangle, and it is thus enough that each $(k-1)$-simplex, i.e., a link, is again incident with an even number of $2$-simplex (two in the figure), having two by two opposite orientation.

\section{A $3$-simplicial complex satisfying conditions \textbf{i}) and \textbf{ii})}
\label{sec:peakedtorus}

The aim of this section is to show how to build a $3$-simplicial complex, $\mathcal{P}$, satisfying conditions \textbf{i}) and \textbf{ii}) for $k=2$ and $k=0$, and thus admitting a synchronous manifold for topological signals defined on faces or nodes. The nodes of this simplicial complex  are placed on a $2$-dimensional square grid with periodic boundary conditions. However each four nodes of any (imaginary) square placquette of this grid are belonging to a distinct tetrahedron.
As it can be directly shown this structure admits an eigenvector $\mathbf{v}$ with elements $|v_i|=1$ for any orientation induced by the node labels, as long as $k\in \{0,2\}$.

 This interesting structure can be obtained by looking at topologies that satisfy conditions \textbf{i}) and \textbf{ii}) for an orientation such that
$\mathbf{v}=\mathbf{u}=(1,\dots,1)^\top$ will lie in $\ker\mathbf{L}_2$ (and of course in $\ker\mathbf{L}_0$). One can show that the resulting oriented simplicial complex has the the same topological properties of the simplicial complex oriented by using the node label order, the reason being that their basis are related by an invertible linear transformation.

For sake of concreteness we will consider the case of topological signals defined on $2$-simplices, belonging to a $3$-simplicial complex. The case $k=0$ being associated to a standard network synchronization phenomenon fits in the standard global synchronisation phenomenon on networks. Conditions \textbf{i)} and \textbf{ii)} translate thus into $\mathbf{B}_2\mathbf{u}=0$ and $\mathbf{B}_3^\top \mathbf{u}=0$. The sought simplicial complex will thus be obtained by assembling together nodes, links, triangles and tetrahedra. Let us notice that once such simplicial complex has been built we show that he can support global synchronization also for topological signals defined on $0$-simplex, i.e., on nodes, while this will not be the case for topological signals defined on $1$-simplex, links, or $3$-simplex, tetrahedra.

\begin{figure*}[ht]
\centering
\includegraphics[scale=0.25]{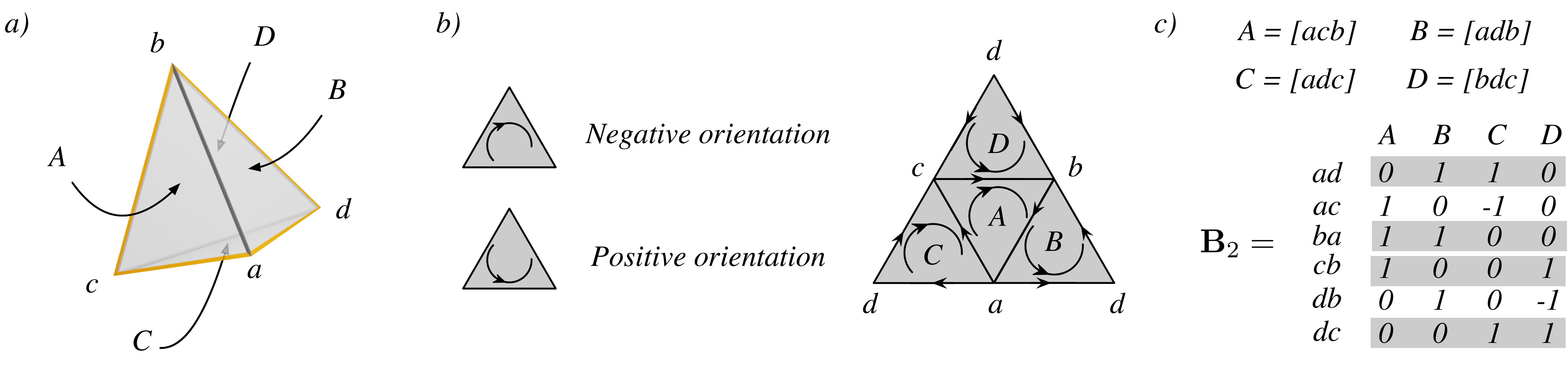}
%\vspace{-1.5cm}
\caption{\textbf{An oriented tetrahedron.} We consider an oriented tetrahedron, i.e., $3$-simplex, both in a $3$D view (panel a)) and in a ``flatten'' $2$D one (panel b)). Given the orientation of the faces and of the links proposed in panel b), we can compute the incidence matrix $\mathbf{B}_2$ (panel c)). We can observe the presence of four links whose contribution to $\mathbf{B}_2\mathbf{u}$ is non zero, they correspond to the grey rows in the matrix and to the orange links in panel a).}
\label{fig:peaked2torus1}
\end{figure*}

Let us start by considering a tetrahedron (see Fig.~\ref{fig:peaked2torus1}a) and let us fix an orientation for its faces, $A$, $B$, $C$ and $D$. For sake of concreteness, we define a face to be positively oriented if one circulates between its vertices in a anticlockwise manner, and negatively oriented on the opposite (see  Fig.~\ref{fig:peaked2torus1}b). To better appreciate the relative orientations of the faces we show the same tetrahedron but ``flattened'', in this way it is clear the presence of two positively oriented ($A$ and $C$) and two negatively oriented ($B$ and $D$) faces, which implies that condition \textbf{i)} is satisfied, i.e., $\mathbf{B}_3^\top \mathbf{u}=0$. 

By labeling the vertices with $a$, $b$, $c$ and $d$, we can assign an orientation to the links (Fig.~\ref{fig:peaked2torus1}b); together with the faces orientation we can compute the incidence matrix $\mathbf{B}_2$ (see Fig.~\ref{fig:peaked2torus1}c). From the latter we can appreciate the existence of four links (rows shaded in light grey in the incidence matrix, also coloured in orange in Fig.~\ref{fig:peaked2torus1}a) that are coherently oriented with two faces, thus in the product $\mathbf{B}_2\mathbf{u}$ they return a $+2$. There are also two links (white rows in the incidence matrix) that are not coherently oriented with the two faces to which they are incident, hence in the product $\mathbf{B}_2\mathbf{u}$ they contribute with a $0$.

\begin{figure*}[ht]
\centering
\includegraphics[scale=0.33]{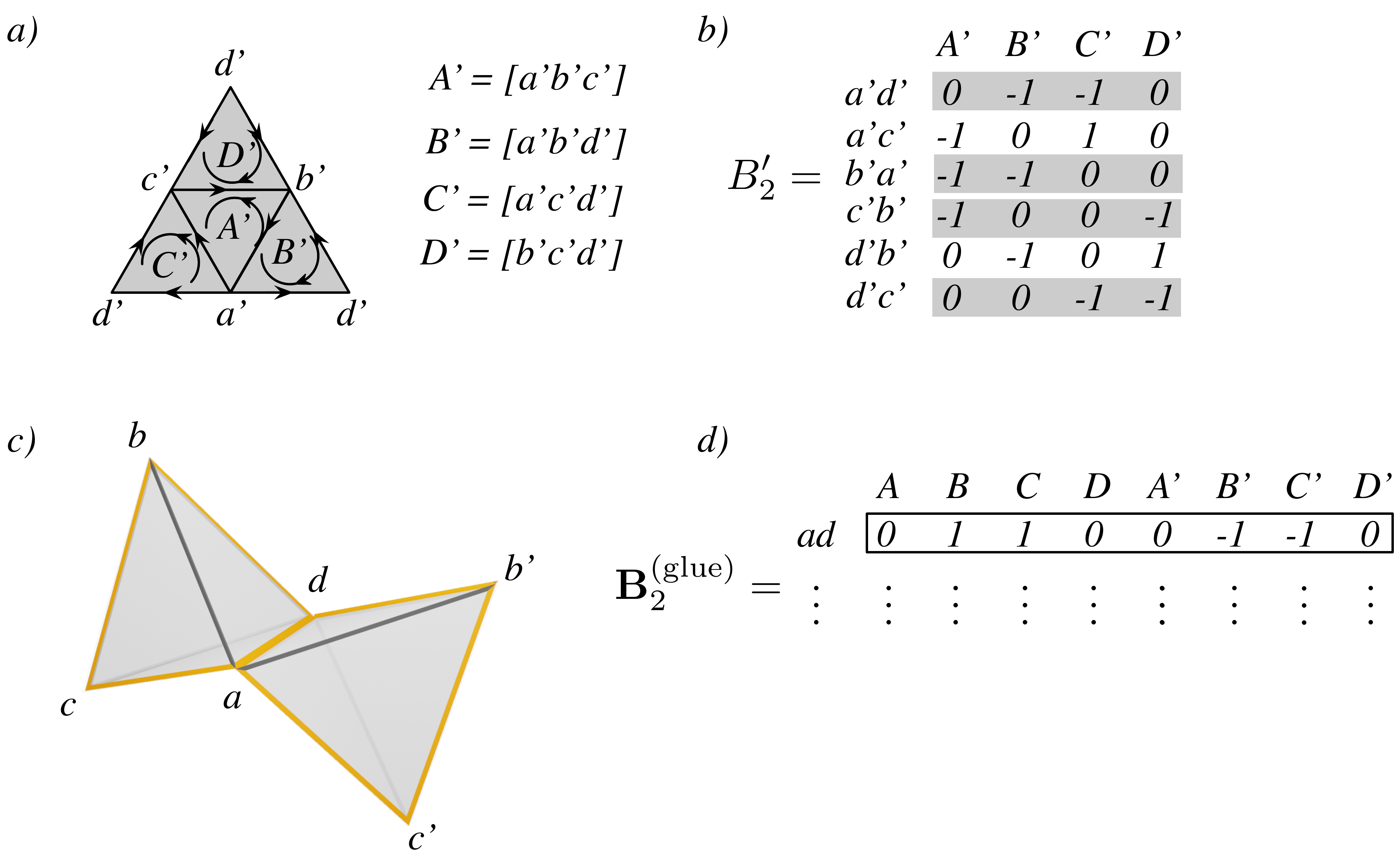}
%\vspace{-1.5cm}
\caption{\textbf{An opposite oriented tetrahedron.} We show a tetrahedron whose faces have an opposite orientation with respect to the ones of Fig.~\ref{fig:peaked2torus1}, while the tetrahedron and its edges are oriented in the same way in both cases (panels a) and b)). In panel c) we show the ``gluing'' process of these two tetrahedra along two selected links, $ad$ and $a'd'$, in such a way the shared link is now incident with four faces and its is coherently oriented with two of them and non coherently with the two other ones (panels c) and d)).}
\label{fig:peaked2torus2}
\end{figure*}

The idea to build the sought $3$-simplicial complex satisfying the constraints $\mathbf{B}_2\mathbf{u}=0$ and $\mathbf{B}_3^\top\mathbf{u}=0$, is to add another oriented tetrahedron with an opposite orientation for its faces with respect to the previous one, but with the same orientation of the links (see Fig.~\ref{fig:peaked2torus2}a). This ``opposite'' tetrahedron will have two positively oriented and two negatively oriented faces and thus it satisfies again $\mathbf{B}_3^\top \mathbf{u}=0$, moreover its incidence matrix $\mathbf{B}_2$ will be the opposite of the previous one (panel b) of Fig.~\ref{fig:peaked2torus2}). This means that we can select an edge in the first tetrahedron (say $ad$) and a second edge in the opposite tetrahedron (say $a'd'$) and ``glue'' them (see Fig.~\ref{fig:peaked2torus2}c). In this way this edge will be incident with four faces, and it will be coherently oriented with two of them and non coherently oriented with the remaining two. Hence the row of the incidence matrix of the ``glued'' tetrahedron, $\mathbf{B}^{(\mathrm{glue})}_2$, will contain two ``$+1$'' and two ``$-1$'', the remaining entries being ``$0$'' (see Fig.~\ref{fig:peaked2torus2}d). In conclusion this link will contribute to $0$ in the product $\mathbf{B}^{(\mathrm{glue})}_2\mathbf{u}$.

\begin{figure*}[ht]
\centering
\includegraphics[scale=0.3]{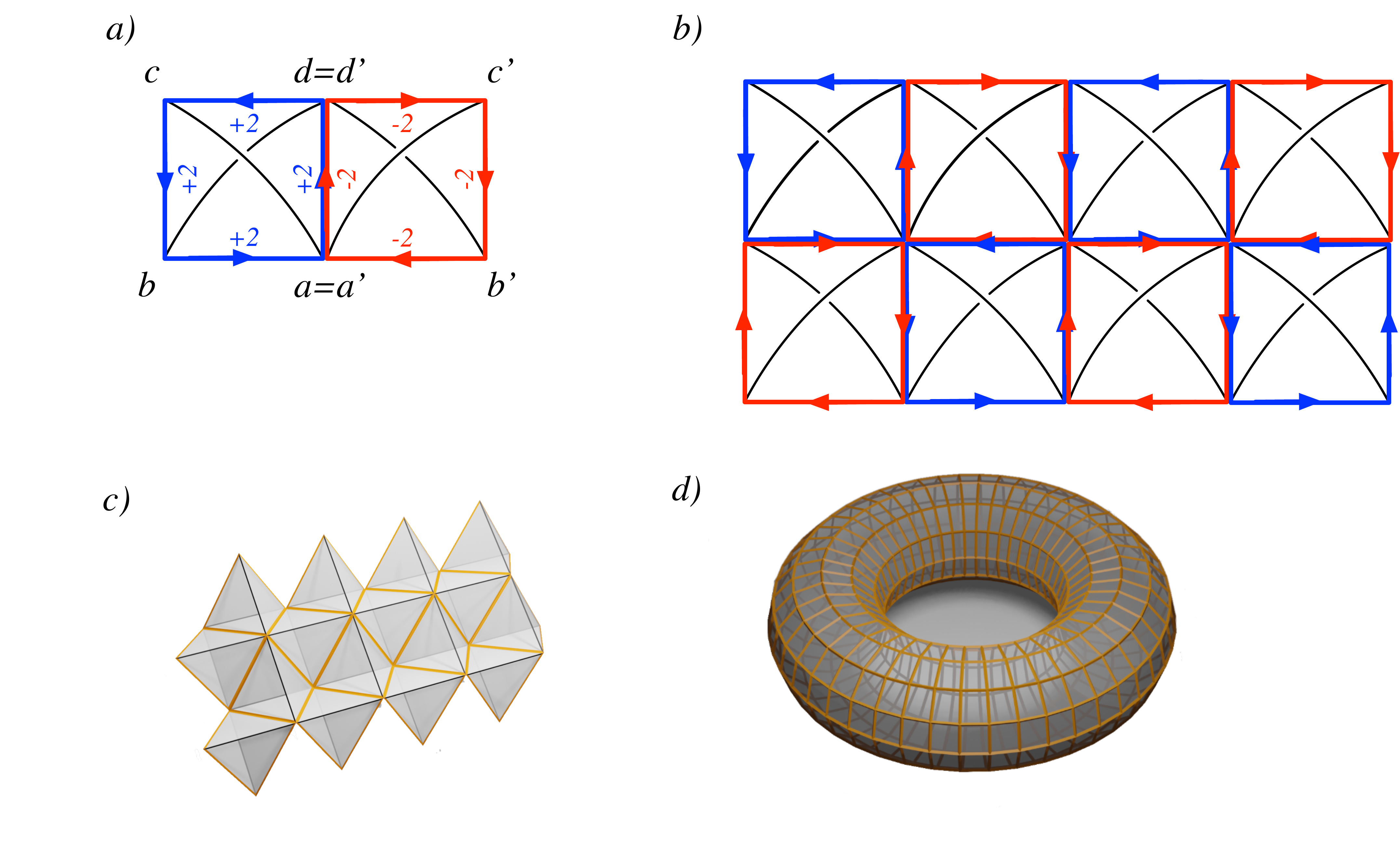}
%\vspace{-1.5cm}
\caption{\textbf{The $3$-simplicial complex $\mathcal{P}$.} We schematically show the construction of the $3$-simplicial complex obtained by gluing together opposite tetrahedra and then by identifying the opposite sides of the square composed by the tetrahedron links contributing with a non zero value to $\mathbf{B}_2\mathbf{u}$ {(the edges in blue and red in panels a-b) or in yellow in panel c)).}}
\label{fig:peaked2torus3}
\end{figure*}

By construction the $3$-simplicial complex obtained by gluing the two opposite tetrahedra contains a link, ($ad$) that is incident with four faces and it is coherently oriented with two of them and incoherently with the other two (see Fig.~\ref{fig:peaked2torus3}a). Moreover such simplicial complex contains six links each one belonging to two faces and coherently oriented with both of them, hence they will contribute with a non zero entry in the product $\mathbf{B}_2\mathbf{u}$. In panel a) of Fig.~\ref{fig:peaked2torus3} they are colored in blue in the case of the first considered tetrahedron and in red for the second one; let us for short say that such links have the ``wrong property". To tackle this issue we repeat the construction show on Fig.~\ref{fig:peaked2torus2} by alternating ``positively'' and ``negatively'' oriented tetrahedra along the two opposite sides of the square obtained by ``flattening'' each tetrahedron and considering only the sides with the ``wrong property'' (see Fig.~\ref{fig:peaked2torus3}b). The resulting structure resemble to a ``waffle'' (see Fig.~\ref{fig:peaked2torus3}c) to which we imposed periodic boundary conditions (see Fig.~\ref{fig:peaked2torus3}c). Eventually taking $L$ squares along the longitudinal direction and $M$ squares along the latitudinal direction, we obtain the sought $3$-simplicial complex that by construction satisfies $\mathbf{B}_2\mathbf{u}=0$ and $\mathbf{B}_3^\top \mathbf{u}=0$. It can thus support a synchronous manifold for topological signals defined on $2$-simplex but not on the $1$-simplex; indeed $\mathbf{B}_2^\top \mathbf{u}\neq 0$, nor the $3$-simplex because $\mathbf{B}_3 \mathbf{u}\neq 0$. Let us observe that in this case $\mathbf{B}_4^\top \mathbf{u}=0$, being the tetrahedron the simplex with the largest dimension and thus $\mathbf{B}_4=0$, however we do not have global synchronization of $3$-topological signals because the used simplex is not a closed manifold without boundary.

\section{A $2$-simplicial complex satisfying conditions \textbf{i}) and \textbf{ii}) for nodes and triangles but not for links}
\label{sec:trianglestorus}

The aim of this section is to build a $2$-simplicial complex satisfying again conditions \textbf{i}) and \textbf{ii}) for $k=2$ and $k=0$. In this case the simplex is formed by triangles connected among them to pave a $2$-torus. We could have used the orientation given by the node labels, but again we decided to introduce an alternative orientation aimed at directly obtaining that $\mathbf{u}=(1,\dots,1)^\top$ lies in $\ker\mathbf{L}_2$. Also in this case this choice does not change the topological properties of the simplicial complex with respect to those of the same simplicial complex oriented by using the node labels orientation.

\begin{figure*}[ht]
\centering
\includegraphics[scale=0.25]{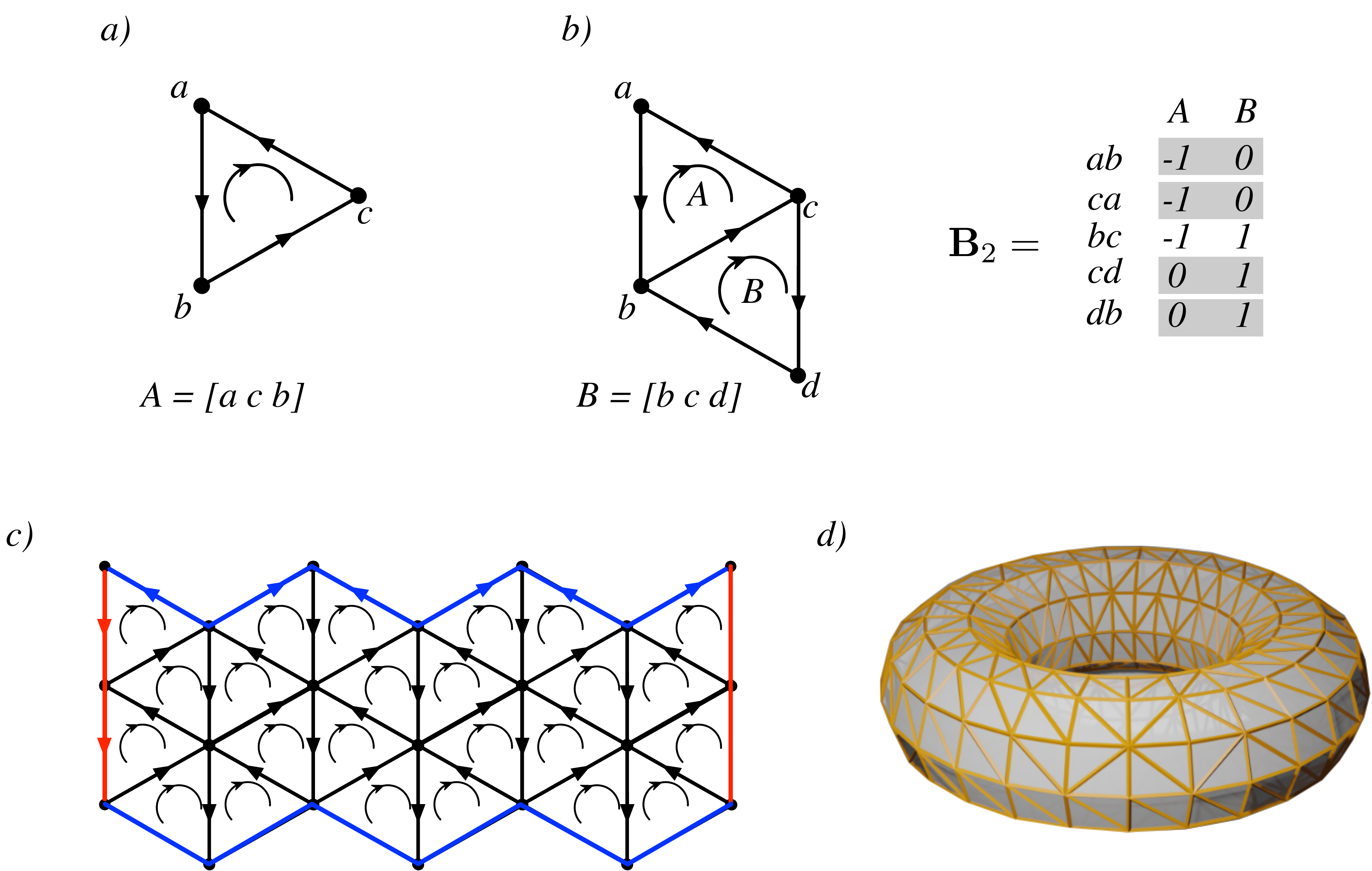}
%\vspace{-1.5cm}
\caption{\textbf{A $2$-torus paved with triangles.} We consider $2$-torus whose surface is paved with oriented triangles (panel a)) organized in such a way that once they share a common link then the latter is coherent with the orientation of one triangle and incoherent with the second; they contribute thus to the matrix $\mathbf{B}_2$ with entries $+1$ and $-1$ (panel b)). Such construction is repeated in the longitudinal and vertical direction and eventually including periodic boundary conditions (panel c)).}
\label{fig:triangtorus1}
\end{figure*}

Consider the case of topological signals defined on $2$-simplices, i.e., the faces of the $2$-simplicial complex. Conditions \textbf{i)} becomes thus $\mathbf{B}_2\mathbf{u}=0$, while condition \textbf{ii)} is automatically satisfied because $\mathbf{B}_3=0$, being the $3$-simplices not allowed. The basic element is thus an oriented triangle, i.e., a $2$-simplex $A=[a\,c\,b]$ whose links are oriented as $[a\, b]$, $[b\, c]$ and $[c\, a]$ (see Fig.~\ref{fig:triangtorus1}a). Consider now a second triangle $B=[b\,c\,d]$ sharing the link $[b\, c]$ with the previous one, and whose links are oriented as $[b\,c]$, $[c\,d]$ and $[d\, b]$. Because of the chosen orientations the shared link determine a $+1$ and a $-1$ entries in the matrix $\mathbf{B}_2$, namely in the product $\mathbf{B}_2\mathbf{u}$ they contribute with a $0$ (see Fig.~\ref{fig:triangtorus1}b). We can then repeat such construction by alternating the two kinds of triangles above defined and eventually identify among them the ``vertical left'' and ``vertical right'' sides (red lines in Fig.~\ref{fig:triangtorus1}c) and also the ``horizontal top'' and ``horizontal bottom'' sides (blue lines in Fig.~\ref{fig:triangtorus1}c). In this way we have obtained a $2$-torus whose surface is paved with triangles and such that by construction we have $\mathbf{B}_2\mathbf{u}=0$.

\section{A $3$-cell complex supporting global synchronization: $3$D-torus}
\label{sec:3Dtorus}

The aim of this appendix is to build a $3$-cell complex made of nodes, links, squares (i.e., $2$-cells) and cubes (i.e., $3$-cells) capable to support global synchronization for topological signals of any dimensions, $k=0,1,2,3$.

\begin{figure*}[ht]
\centering
\includegraphics[scale=0.25]{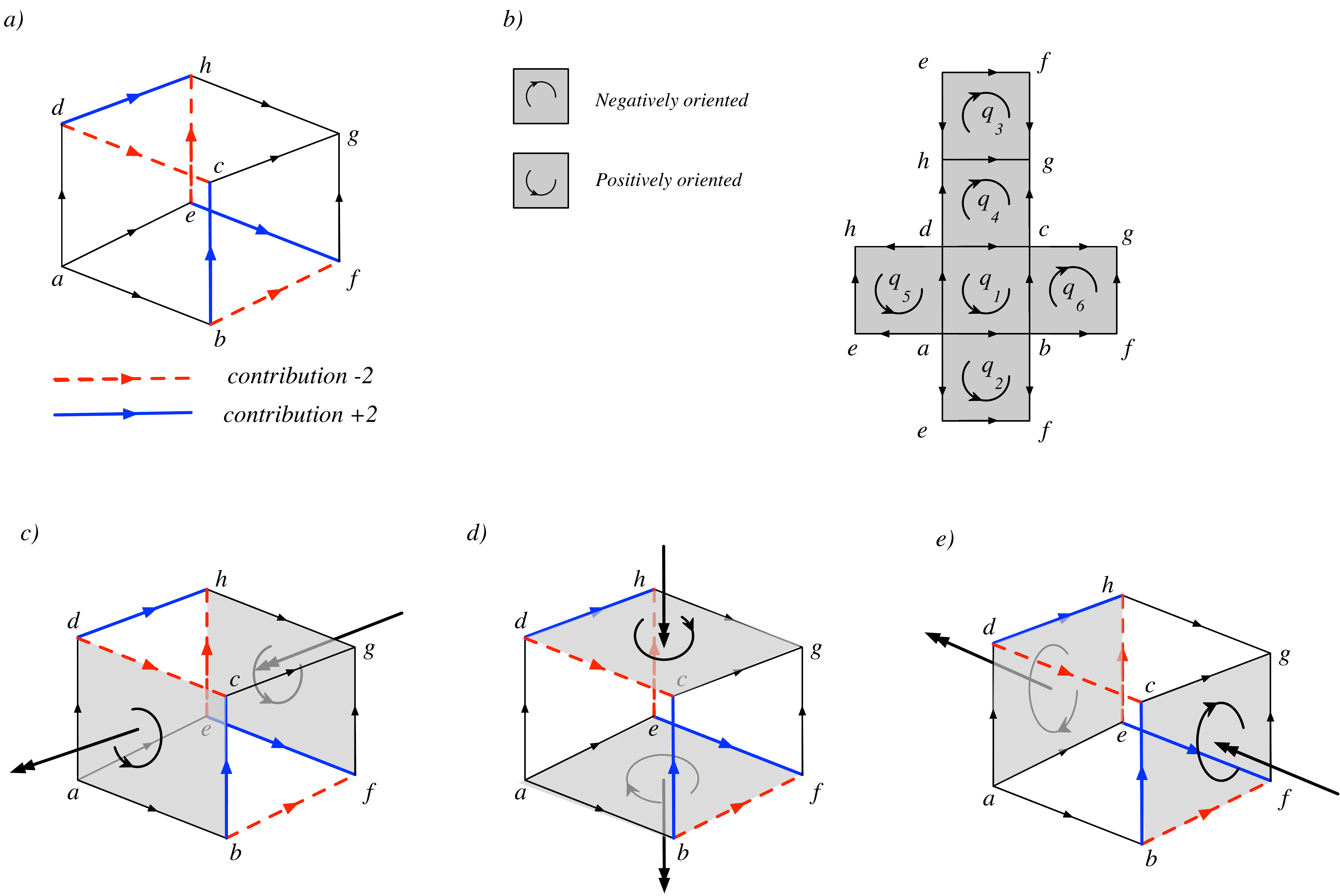}
\caption{\textbf{An oriented cube.} We show an oriented cube, both in a $3$D view (panel a)) then in a ``flatten'' $2$D one (panel b)). Given an orientation of the faces and of the links (see panels a) and b)), we can compute the incidence matrix $\mathbf{B}_2$ and identify the link whose contribution to $\mathbf{B}_2\mathbf{u}$ is not zero. This allows to emphasise $6$ links forming a closed loop (thicker dashed red and solid blue lines in panel a)) in the cube. Let us observe that opposite faces of the cube must be incoherently oriented (see panels c)-d) and e)).}
\label{fig:3DtorusA}
\end{figure*}

Let us consider a cube formed by six $2$-cells, the squares (see Fig.~\ref{fig:3DtorusA}a), and we can orient the faces to have three positive and three negative contributions to $\mathbf{B}_3^\top \mathbf{u}$ is in such a way the latter sums to zero (see panels b)). It is important for the following construction to incoherently orient opposite faces (see panels c)-d) and e)), this is because once we will ``glue'' together several cubes, the shared faces should have the right orientation to ensure $\mathbf{B}_2\mathbf{B}_3=0$ . 

We then orient the links according to a left-right on the horizontal  direction, bottom-up in the vertical one and front-back in the transversal direction (see again panels a) and b)). Once faces and links have been oriented we can compute the matrices $\mathbf{B}_k$, $k=1,\dots,3$ and we can observe that six links exist, three of which contributing with ``$+2$'' (solid blue line) or ``$-2$'' (dashed red line) to $\mathbf{B}_2 \mathbf{u}$.

\begin{figure*}[ht]
\centering
\includegraphics[scale=0.285]{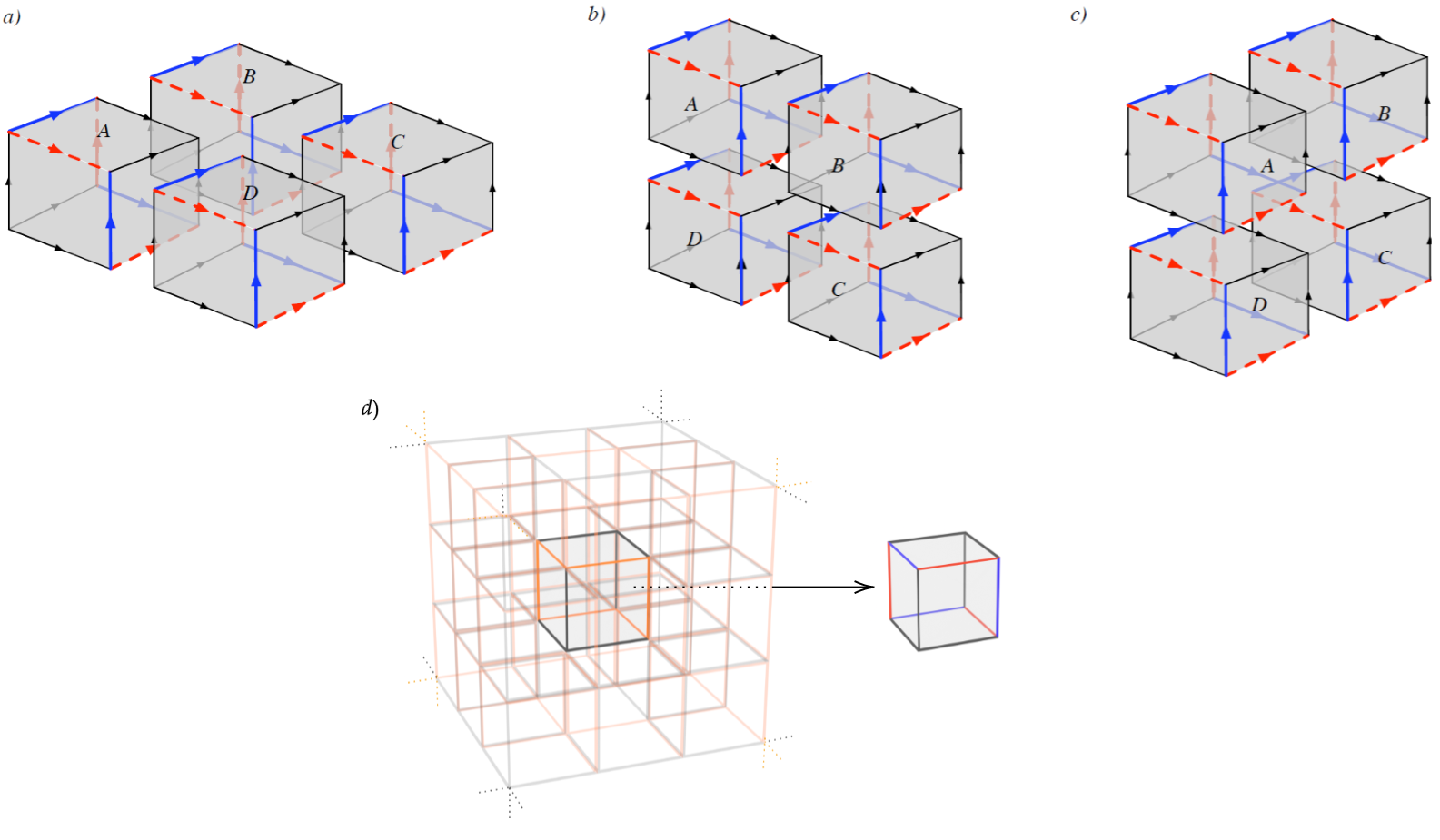}
\caption{\textbf{Gluing together four oriented cubes along the three dimensions.} We show how to glue together four cubes in the horizontal (panel a)), vertical (panel b)) and vertical transversal (panel c)) plane. We can observe that in all cases the link common to eight faces (and four cubes) returns a null contribution to $\mathbf{B}_1\mathbf{u}$, indeed it contributes with $0$ to two couples of faces (thin black line), with $+2$ to a couple of faces (blue thick line) and with $-2$ to the remaining couple of faces (dashed red line). {The final construction is schematically depicted in panel d). The cube is repeated along each one of the three spatial dimensions and periodic boundary conditions are implemented connecting opposite faces of the cubic array  (represented with a lower opacity). The basic element of the construction is shown in the centre. As in the other figures, the non zero contributing edges are shown in orange and in dark grey all the others. {The edges coloured with the same color index as in panels a)-c) are also shown for orientation reference.}}}
\label{fig:3DtorusB}
\end{figure*}

Once we have obtained the oriented cube, we can ``glue'' together several copies of the same cube along the three directions (see Fig.~\ref{fig:3DtorusB}). Because of the orientations given to the faces (see Fig.~\ref{fig:3DtorusA}), the face shared by each couple of cubes will have the right orientation. Moreover this gluing process ensures also that $\mathbf{B}_2\mathbf{u}=0$; indeed as it can been observed from Fig.~\ref{fig:3DtorusB} the unique link incident with four cubes, hence eight faces, has a zero contribution to $\mathbf{B}_2\mathbf{u}$. By considering for instance the configuration shown in panel a), the vertical link common to the four cubes A, B, C and D will contribute with $+2$ (it has been coloured in blue according to the scheme presented in Fig.~\ref{fig:3DtorusA}) because of the two faces of the B cube, with $-2$ (it has been coloured in red according to the scheme presented in Fig.~\ref{fig:3DtorusA}) because of the two faces of the D cube and $0$ (it has been coloured in black according to the scheme presented in Fig.~\ref{fig:3DtorusA}) to the faces in the cubes A and C. The same analysis can be done for the constructions in the vertical direction (panel b)) and in the transversal one (panel c)).
Finally by imposing periodic boundary conditions in the three directions we obtain a $3$D torus with the required properties. Let us observe that the links orientation together with the periodic boundary conditions ensure $\mathbf{B}_1\mathbf{u}=0$. 

\section{Analysis of the Stuart-Landau model}
\label{sec:SL}

The aim of this section is to provide the interested reader more details about the global synchronization of identical Stuart-Landau systems defined on top of simplicial and cell complexes introduced in the main text. 

Let us thus assume to deal with a $k$ dimensional topological signal associated to a nonlinear SL oscillator, whose time evolution is given by
\begin{equation*}
 \frac{dw}{dt} = \sigma w -\beta w|w|^2\, ,
\end{equation*}
where $\sigma=\sigma_\Re+i\sigma_\Im$ and $\beta=\beta_\Re+i\beta_\Im$ are complex model parameters. One can easily prove that the former system admits a limit cycle solution $\hat{z}(t)=\sqrt{\sigma_\Re/\beta_\Re}e^{i\omega t}$, where $\omega=\sigma_\Im-\beta_\Im \sigma_\Re/\beta_\Re$, is the signal frequency. Such solution is stable provided $\sigma_\Re>0$ and $\beta_\Re>0$, conditions that we hereby assume.

We now consider $N_k$ SL topological oscillators and we couple them using the the $(k-1)$ and $(k+1)$-faces of a simplicial complex. Let also assume the coupling to be given by the nonlinear function $h(w)=\mu w |w|^{m-1}$, where $m$ is a positive integer and $\mu=\mu_\Re+i\mu_\Im$ is a complex parameter defining the interaction strength. Observe that such coupling has been recently introduced and studied in the framework on synchronization on time varying networks in~\cite{CarlettiFanelli2022}. In conclusion we are interested in studying the system
\begin{equation}
\frac{dw_i}{dt} = \sigma w_i -\beta w_i|w_i|^2+\mu \sum_{j=1}^{N_k} L_k(i,j) w_j |w_j|^{m-1}\quad \forall i=1,\dots, N_k\, .
\label{eq:coupledSL}
\end{equation}

We are now interested in studying the stability of the reference limit cycle solution $\hat{z}(t)$. If this condition is realised, then the system shows global synchronization. To achieve this goal we introduce real ``small'' functions $\rho_j(t)$ and $\theta_j(t)$ and rewrite $w_j(t)$ as follows 
\begin{equation}
\label{eq:wpert}
w_j(t)=\hat{z}(t)(1+\rho_j(t))e^{i\theta_j(t)}\, ,
\end{equation}
{where $\rho_j(t)$ and $\theta_j(t)$ are real valued functions.}
We now insert the previous expression in the coupled Eq.~\eqref{eq:coupledSL}.
%\begin{equation}
%\label{eq:coupledSL}
%\frac{dw_i}{dt} = \sigma w_i -\beta w_i|w_i|^2-\mu %\sum_{j=1}^{N_k} L_k(i,j) w_j |w_j|^{m-1}\, ,
%\end{equation} valid for every $i=1,\dots, N_k$. 
By using the expression for $\hat{z}(t)$ and by expanding the resulting equation up to the first order in $\rho_j$ and $\theta_j$, we eventually obtain
\begin{equation}
\label{eq:maineqSLlin}
\begin{dcases}
\frac{d\rho_j}{dt} = -2\sigma_\Re \rho_j  -\left(\frac{\sigma_\Re}{\beta_{\Re}}\right)^{\frac{m-1}{2}}\sum_{\ell=1}^{N_k} L_k(j,\ell)\left(m\mu_\Re \rho_\ell-\mu_\Im \theta_\ell\right)\\
\frac{d\theta_j}{dt} = -2\beta_\Im\frac{\sigma_\Re}{\beta_{\Re}} \rho_j  -\left(\frac{\sigma_\Re}{\beta_{\Re}}\right)^{\frac{m-1}{2}}\sum_{\ell=1}^{N_k} L_k(j,\ell)\left(m\mu_\Im \rho_\ell+\mu_\Re \theta_\ell\right)\, .
\end{dcases}
\end{equation}

We can then decompose $\rho_j(t)$ and $\theta_j(t)$ on the orthonormal eigenbasis $\pmb{\phi}_k^{(\alpha)}$, $\alpha=1,\dots,N_k$, of the Laplace matrix $\mathbf{L}_k$:
\begin{equation}
\label{eq:projection}
 \rho_j=\sum_\alpha \hat{\rho}_\alpha \phi^{(\alpha)}_k(j) \text{ and } \theta_j=\sum_\alpha \hat{\theta}_\alpha \phi^{(\alpha)}_k(j) \, ,
\end{equation}
to eventually obtain
\begin{equation}
\label{eq:maineqSLlinMSFApp}
\begin{dcases}
\frac{d{\hat{\rho}}_\alpha}{dt} = -2\sigma_\Re \hat{\rho}_\alpha  -\left(\frac{\sigma_\Re}{\beta_{\Re}}\right)^{\frac{m-1}{2}}\Lambda^{(\alpha)}_k\left(m\mu_\Re \hat{\rho}_\alpha-\mu_\Im \hat{\theta}_\alpha\right)\\
\frac{d{\hat{\theta}}_\alpha}{dt} = -2\beta_\Im\frac{\sigma_\Re}{\beta_{\Re}} \hat{\rho}_\alpha  -\left(\frac{\sigma_\Re}{\beta_{\Re}}\right)^{\frac{m-1}{2}}\Lambda^{(\alpha)}_k\left(m\mu_\Im \hat{\rho}_\alpha+\mu_\Re \hat{\theta}_\alpha\right)\, .
\end{dcases}
\end{equation}

Let us observe that Eq.~\eqref{eq:maineqSLlinMSFApp} is autonomous, hence one can compute its eigenvalues and define the largest real part of the latter ones, say $\lambda$, named in the literature {\em dispersion relation}. One can thus conclude that if $\lambda <0$ the reference solution is stable and hence the system globally synchronizes. The same reasoning can be done whenever the generic MSF~\eqref{eq:coupledlinspect} is autonomous. 

Let us also observe that a similar conclusion can be obtained if $\mathbf{s}(t)$ is a periodic solution by resorting to Floquet analysis; calling again $\lambda$ the largest real part of the Floquet eigenvalues we can show that if $\lambda<0$ then the reference solution is stable and the system globally synchronizes. In the general case, one has to (numerically) compute the Lyapunov exponent of~\eqref{eq:coupledlinspect} and infer about the stability of the reference solution using the Lyapunov theory. Let us observe that to stress the dependence on the simplex eigenvalues we will also write $\lambda_\alpha=\lambda(\Lambda_k^{(\alpha)})$.

Back to Eq.~\eqref{eq:maineqSLlinMSFApp} one can infer the stability of the reference solution and thus of the global simplicial synchronization by studying if the perturbations $\rho_j$ and $\theta_j$ fade away, or equivalently if their projections $\hat{\rho}_\alpha$ and $\hat{\theta}_\alpha$ vanish. Sufficient conditions are obtained by assuming an exponential behavior, namely {$\hat{\rho}_\alpha \sim \tilde{\rho}_{\alpha} e^{\lambda_\alpha t}$ and $\hat{\theta}_\alpha \sim \tilde{\theta}_{\alpha} e^{\lambda_\alpha t}$ with time-independent $\tilde{\rho}_{\alpha}$ and $\tilde{\theta}_{\alpha}$}. Inserting this ansatz into~\eqref{eq:maineqSLlinMSFApp} and {imposing that $(\tilde{\rho}_{\alpha},\tilde{\theta}_{\alpha})\neq (0,0)$,} one gets the following equation for $\lambda_\alpha$
\begin{equation}
\label{eq:disprel}
\lambda_\alpha^2+\lambda_\alpha \left( \left(\frac{\sigma_\Re}{\beta_{\Re}}\right)^{\frac{m-1}{2}} \mu_\Re \Lambda^{(\alpha)}_k(m+1)+2\sigma_\Re\right)+m\left(\frac{\sigma_\Re}{\beta_{\Re}}\right)^{{m-1}}\left(\Lambda^{(\alpha)}_k\right)^2(\mu_\Re^2+\mu_\Im^2)+2\Lambda^{(\alpha)}_k\left(\frac{\sigma_\Re}{\beta_{\Re}}\right)^{\frac{m-1}{2}}\left(\mu_\Re\sigma_\Re+\mu_\Im \beta_\Im \frac{\sigma_\Re}{\beta_\Re}\right)=0\, .
\end{equation}
We eventually define the {\em dispersion relation} (or maximum Floquet exponent) $\lambda = \max_\alpha \Re\lambda_\alpha$. Let us observe that $\lambda_1=0$, that is $\lambda$ vanishes if evaluated on $\Lambda^{(1)}_k=0$; this is because the reference solution is a limit cycle. By considering then the behavior of $\lambda_\alpha$ for $\Lambda^{(\alpha)}_k$ close to zero, one can develop the root $\lambda_\alpha$ as follows
\begin{eqnarray*}
\lambda_\alpha &\sim& \frac{1}{2}\left[ -\left(\frac{\sigma_\Re}{\beta_\Re}\right)\mu_\Re \Lambda^{(\alpha)}_k(m+1)-2\sigma_\Re +2\sigma_\Re \left(1+ \frac{1}{2}\left(\frac{\sigma_\Re}{\beta_\Re}\right)^{(m-1)/2} \Lambda^{(\alpha)}_k \left(\frac{\mu_\Re}{\sigma_\Re}(m-1) -2 \frac{\beta_\Im \mu_\Im}{\beta_\Re\sigma_\Re}\right)\right)+\dots \right]\\
&=&-\Lambda^{(\alpha)}_k \left(\frac{\sigma_\Re}{\beta_\Re}\right)^{(m-1)/2} \left(\mu_\Re +\frac{\beta_\Im \mu_\Im}{\beta_\Re}\right)+\dots\, .
\end{eqnarray*}

Hence there exists an interval of values for $\Lambda^{(\alpha)}_k$ such that $\lambda_\alpha>0$, namely the global synchronization cannot be achieved, if and only if
\begin{equation*}
 \mu_\Re +\mu_\Im \frac{\beta_\Im}{\beta_\Re}<0\, .
\end{equation*}

\begin{figure}[ht]
\centering
\includegraphics[scale=0.28]{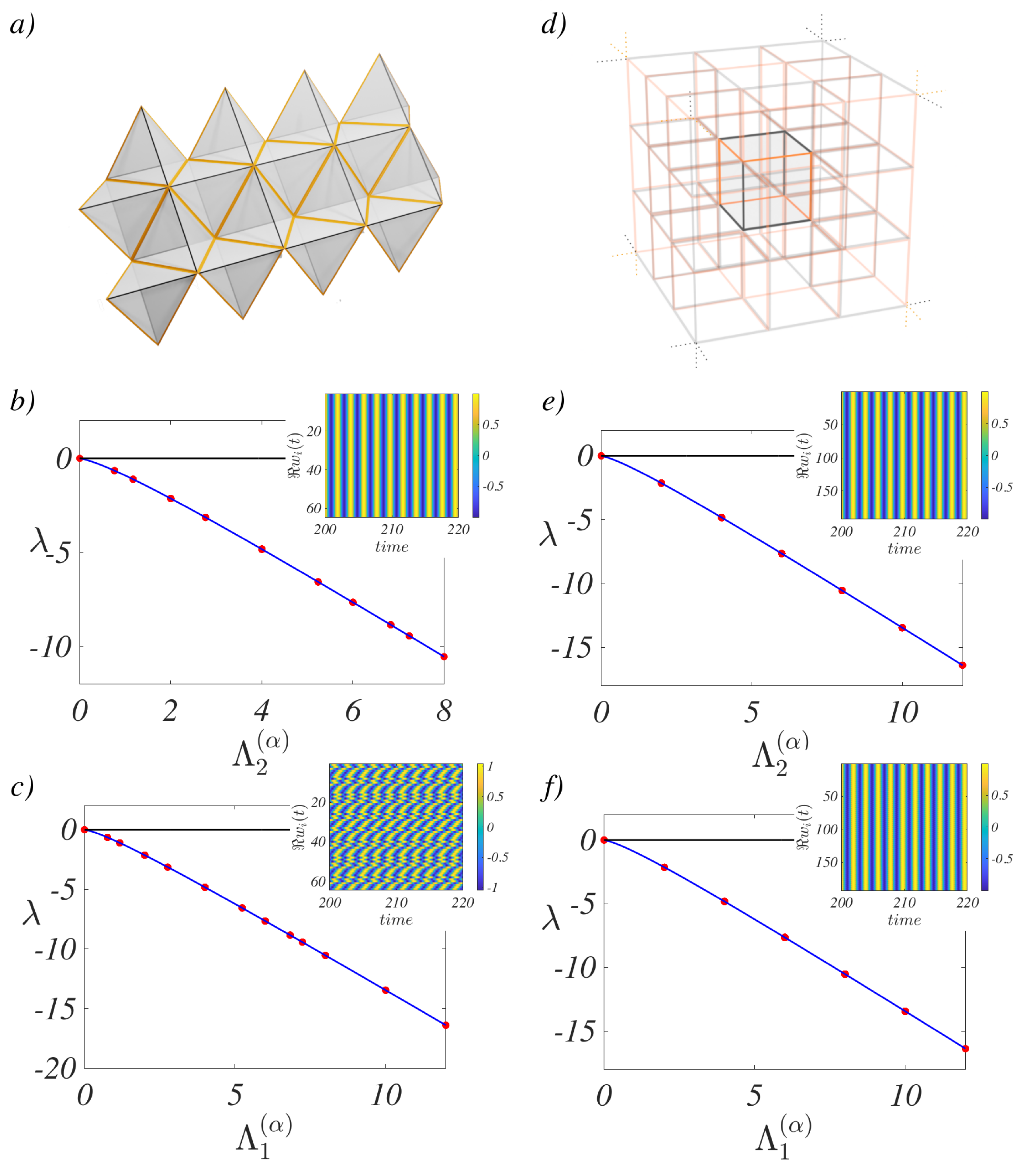}
%\vspace{-1.5cm}
\caption{\textbf{Dispersion relation for topological Stuart-Landau model.} The left panels refer to the $3$-simplicial complex schematically represented in panel a). In panel b) we report the dispersion relation $\lambda$ as a function of the eigenvalues $\Lambda^{(\alpha)}_2$  in the case of topological complex amplitudes defined on $2$-faces, i.e., triangles, while as a function of $\Lambda^{(\alpha)}_1$ in panel c) dealing with signals defined on links. Similar functions are reported in the right panels in the case of $3$-cell complex schematically represented in panel d). The model parameters have been fixed to some generic values, $\sigma = 1.0+4.3 i$, $\beta = 1.0+1.1 i$, $\mu = 1.0-0.5i$ and $m = 3$, and are the same used to obtain the results reported in Fig.~\ref{fig:SLsmplx}.}
\label{fig:SLreldisp}
\end{figure}

The above presented theory is applied to topological SL signals defined on top of the $3$-simplicial complex and the $3$-cell complex previously introduced. The model parameters have been set to some generic values allowing for a negative dispersion relation (see Fig.~\ref{fig:SLreldisp}(b-c) for the simplicial complex and  Fig.~\ref{fig:SLreldisp}(e-f) for the cell complex). However once the complex amplitudes are defined on $2$-faces, i.e., triangles or squares, the system globally synchronizes as we can appreciate from the inset in panel b) and e), while SL oscillators defined on links have a different behavior if we are dealing with a simplicial complex where they cannot globally synchronize (see panel c)) or a cell complex where global synchronization is achieved (see panel f)). Those different behaviors result from the fact that $\mathbf{u}\in\ker\mathbf{L}_1$ for the cell complex while $\mathbf{u}\not\in\ker\mathbf{L}_1$ for the simplicial complex, in the latter case synchronization cannot be achieved because of the presence of the topological obstruction.

\begin{figure}[ht]
\centering
\includegraphics[scale=0.28]{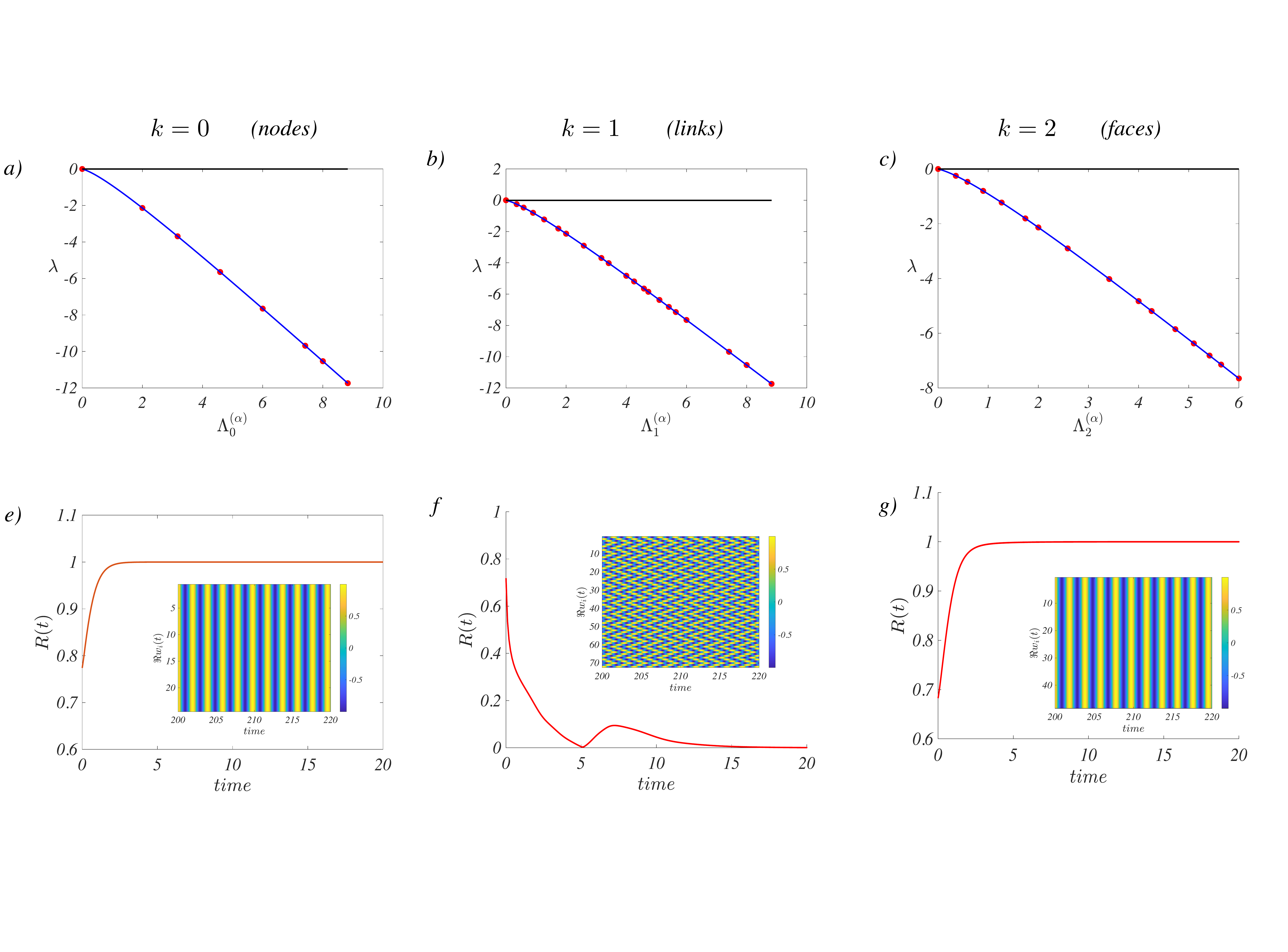}
\vspace{-1.5cm}
\caption{\textbf{Dispersion relation for topological Stuart-Landau model (II).} We consider the $2$-torus paved with triangles. The left panel refers to topological signals defined on nodes, i.e., $0$-simplices, the middle panel to the case of links, i.e., $1$-simplices and the right panel to the case of faces, i.e., $2$-simplices panels. Top panels show the dispersion relation and we can observe that in all cases the latter is negative except for the zero value associated to $\Lambda^{(1)}_k=0$, $k=0,1,2$. Bottom panels present the (generalized) order parameters while the inset report the real part of the signals. The model parameters have been fixed to some generic values, $\sigma = 1.0+4.3 i$, $\beta = 1.0+1.1 i$, $\mu = 1.0-0.5i$ and $m = 3$, and are the same used to obtain the results reported in Fig.~\ref{fig:SLsmplx}.}
\label{fig:SLreldispX}
\end{figure}

Let us conclude this section by showing the dispersion relation for the topological SL signals defined on the $2$-torus paved with triangles. By assuming the same model parameters as in Fig.~\ref{fig:SLreldisp} the numerical results presented in Fig.~\ref{fig:SLreldispX} confirm our theory, global synchronization can be achieved only for topological signals defined on nodes, i.e., $k=0$ simplices (see left panel), or triangles, i.e., $k=2$ simplices (see right panel). Indeed we have $\mathbf{u}\in\ker \mathbf{L}_0$ and $\mathbf{u}\in\ker \mathbf{L}_2$; on the other hand $\mathbf{u}\not\in\ker \mathbf{L}_1$ and thus topological signals defined on links cannot globally synchronize (see middle panel).

\end{document}